\pdfoutput=1
\documentclass[runningheads]{llncs}
\usepackage[T1]{fontenc}
\usepackage{graphicx}
\usepackage{booktabs}
\usepackage{multirow}
\usepackage[misc]{ifsym}
\newcommand{\corr}{(\Letter)}
\usepackage{amsmath,amssymb}
\usepackage{url}
\usepackage{tikz}
\usetikzlibrary{positioning,shapes.geometric,arrows.meta,fit,backgrounds,calc}

\begin{document}

\title{From Rules to Neural Graphs: Scalable Structured Prediction for Patent Prior Art Search}

\titlerunning{From Rules to Neural Graphs for Patent Prior Art Search}

\author{Nikolai Zenovkin\inst{1}\orcidID{0009-0007-6283-427X}\corr \and
Sebastian Bj\"orkqvist\inst{1}\orcidID{0009-0006-9039-8623}}
\authorrunning{N. Zenovkin and S. Bj\"orkqvist}
\institute{IPRally Technologies Oy, Helsinki, Finland\\
\email{\{nikolai.zenovkin,sebastian\}@iprally.com}}

\maketitle

{\let\thefootnote\relax\footnotetext{This preprint has not undergone peer review or any
post-submission improvements or corrections. The Version of Record of this contribution is
published in \emph{Machine Learning and Knowledge Discovery in Databases. Applied Data Science
Track} (ECML PKDD 2026), LNAI 16949, and is available online at
\url{https://doi.org/10.1007/978-3-032-37682-4_31}.}}

\begin{abstract}
Patent search requires processing documents routinely exceeding tens of thousands of tokens. Most neural retrieval approaches operate on truncated inputs, limiting their effectiveness. Graph-based retrieval addresses this by representing each patent as a structured \textit{invention graph}, but constructing these graphs relies on brittle rule-based parsers. We present the \textit{neural parser}, which adapts biaffine attention from dependency parsing to predict invention graphs directly from patent text. Our \textit{local biaffine attention} restricts pairwise scoring to a sliding window, reducing complexity from $O(n^2)$ to $O(n \cdot w)$. Since local and global scoring share the same weights, the model trains on short sequences and deploys on documents exceeding 40,000 tokens without retraining. Distilled from 1 million rule-parsed documents, it surpasses its teacher at 3$\times$ lower inference cost: neural graphs improve citation recall by 0.5\% on short queries and 1.1\% on full documents in a downstream Graph Transformer retrieval system.

\keywords{Patent search \and Prior art search \and Information retrieval \and Dense retrieval \and Structured prediction \and Long document processing \and Biaffine attention \and Knowledge distillation \and Natural language processing}
\end{abstract}

\section{Introduction}
\label{sec:introduction}

Patent prior art search---identifying existing publications that may anticipate or render obvious a patent application's claims---is essential in intellectual property law. Traditional Boolean keyword search~\cite{clarke2018basics,dhondt2010clef} requires substantial domain expertise and multiple iterations, as similar concepts may use entirely different terminology across patent documents.

\paragraph{Dense Retrieval for Patents.}
Neural dense retrieval models overcome terminology variation by learning semantic representations~\cite{ghosh2024paecter,acikalin2022patent,freunek2021bert}, but face a fundamental length constraint: patent documents are extremely long. In our training corpus of 1 million patent documents, descriptions average 55{,}000 characters (median: 40{,}000), with the 90th percentile at 104{,}000 characters. Current models process only truncated inputs---typically titles and abstracts~\cite{ghosh2024paecter,acikalin2022patent,freunek2021bert}---sacrificing coverage of claim-relevant content in the full text.

\paragraph{Graph-Based Compression.}
Invention graphs address this by extracting core technical features and organizing them hierarchically~\cite{bjorkqvist2023building}, compressing patents into structured representations suitable for neural processing. Graph Transformers~\cite{daniell2025efficient} process these structures via sparse attention over graph edges, enabling effective handling of documents exceeding 40,000 tokens---far beyond what standard dense retrieval supports~\cite{zhu2024longembed}.

\begin{figure}[t]
\centering
\includegraphics[width=\linewidth]{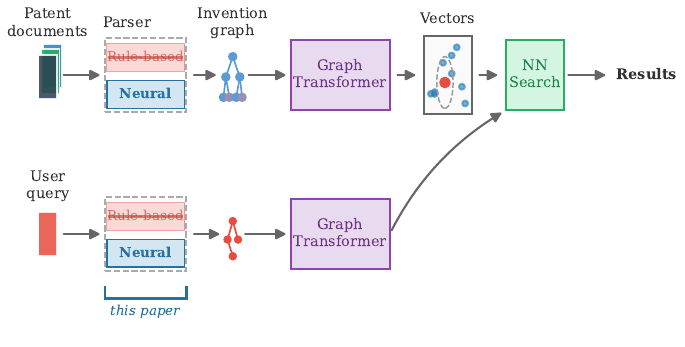}
\caption{End-to-end patent prior art search pipeline. Patent documents and user queries are parsed into invention graphs, encoded by a Graph Transformer~\cite{daniell2025efficient} into dense vectors, and matched via nearest neighbor search. This work replaces the rule-based parser with a neural alternative (highlighted).}
\label{fig:pipeline}
\end{figure}

\paragraph{The Rule-Based Bottleneck.}
Existing graph-based patent search systems~\cite{bjorkqvist2023building,daniell2025efficient} rely entirely on rule-based parsers with extensive hand-crafted patterns. While achieving high precision on matched patterns, these parsers suffer from brittleness: they fail silently on novel linguistic constructions and require continuous manual curation as patent language evolves. A natural alternative is a learned graph induction model, but standard neural structured prediction methods---such as biaffine dependency parsers~\cite{dozat2017deep}---require scoring all $O(n^2)$ token pairs, making them intractable for documents with tens of thousands of tokens. Obtaining labeled training data at corpus scale is a further obstacle: manually annotating graph structures across millions of documents is impractical.

We address both challenges with the following contributions:
\begin{itemize}
\item \textbf{Scalable structured prediction}: We extend sliding-window biaffine attention~\cite{wang2024docban} from flat span extraction to structured prediction (tree/DAG induction) over full-length patent documents. Crucially, local and global scoring share the same weight matrix, so the model trains on short sequences with global attention and deploys on long documents with local attention---no retraining required.
\item \textbf{Student surpasses teacher}: Via knowledge distillation on 1 million rule-parsed documents, our neural parser generalizes beyond hand-crafted patterns, yielding better graph quality and improved downstream prior art search, while offering 3$\times$ lower inference cost and flexibility for diverse data sources (human annotation, LLM-generated labels).
\end{itemize}

\noindent To support reproducibility, we release: (1)~the evaluation set of 5{,}000 annotated patent documents used for intrinsic evaluation, (2)~a 5{,}000-document sample from the training set, and (3)~the evaluation framework for computing the graph quality metrics reported in this paper.\footnote{Available at \url{https://github.com/iprally/invention_graph_eval}.}

\section{Related Work}
\label{sec:background}

\subsection{Patent Retrieval and Graph Representations}
\label{sec:patent-retrieval}

Dense retrieval has become the leading neural approach to patent search. Transformer models fine-tuned for patents~\cite{acikalin2022patent,freunek2021bert}, including PaECTER~\cite{ghosh2024paecter} with citation-informed embeddings, outperform general-purpose models~\cite{reimers2019sentence}, but even recent benchmarks support at most 4,096 tokens~\cite{merrouni2025patenteb}---far short of typical patent descriptions. LongEmbed~\cite{zhu2024longembed} confirms that embedding models degrade substantially beyond 2,048 tokens.

Graph-based retrieval addresses this. Invention graphs encoded with Tree-LSTMs~\cite{bjorkqvist2023building,tai2015improved} significantly improve retrieval over bag-of-words baselines, and a Graph Transformer approach~\cite{daniell2025efficient,dwivedi2021generalization,ying2021transformers} achieves further gains via sparse attention over graph edges. Both systems rely on rule-based parsers---the bottleneck this work addresses. Open relation extraction from patent claims~\cite{geng2021open} produces flat triples rather than the hierarchical representations required for structured retrieval.

\subsection{Neural Structured Prediction for Long Documents}
\label{sec:neural-parsing}

Our formulation adapts biaffine attention from dependency parsing~\cite{dozat2017deep}, where separate MLPs project words into head and dependent roles and a biaffine function scores pairwise arcs. While biaffine parsing remains the dominant paradigm for syntactic and semantic dependency parsing, existing methods operate at sentence level. DocBAN~\cite{wang2024docban} recently applied sliding-window biaffine attention to document-level NER, reducing quadratic complexity to linear for span extraction. Our work addresses the harder problem of structured prediction---inducing tree and DAG structures via per-word parent selection rather than identifying flat spans---at much longer document scales (40,000+ vs.\ thousands of tokens).

Processing such long sequences requires efficient attention. Longformer~\cite{beltagy2020longformer} and BigBird~\cite{zaheer2020bigbird} introduced sparse attention combining local windows with global tokens in the transformer encoder. We use Longformer as our encoder but introduce sparsity in the biaffine scoring layer---the computational bottleneck for structured prediction. A key distinction from DocBAN, which adds a U-Net module for cross-window context, is that our local biaffine mechanism uses exactly the same weight matrix as global scoring. This weight-sharing property enables training with global attention on short sequences and deploying with local attention on long documents without retraining (Section~\ref{sec:local-attention}).

We train via knowledge distillation~\cite{hinton2015distilling} from a rule-based teacher. Combined with loss weighting (Section~\ref{sec:training-objective}), this provides sufficient signal for the model to generalize beyond hand-crafted patterns without manual annotation at corpus scale.

\section{Methodology}
\label{sec:methodology}

\subsection{Invention Graphs}
\label{sec:problem-formulation}

\begin{figure}[t]
\centering
\includegraphics[width=\linewidth]{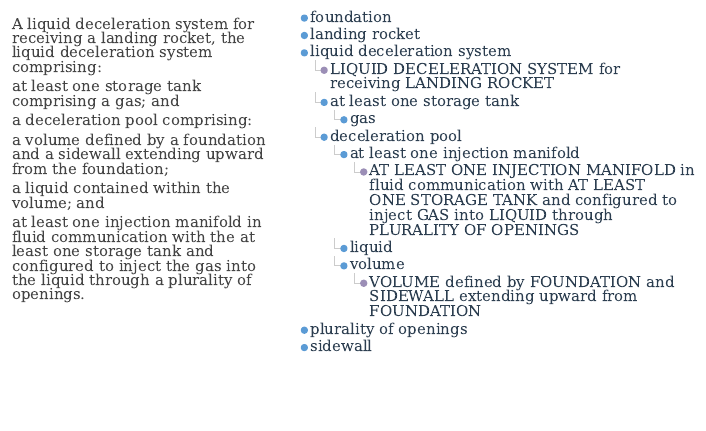}
\caption{A patent claim (US20250197034A1) and the corresponding invention graph. Blue bullets denote feature nodes; purple bullets denote functional relations. Indentation reflects hierarchical (part-of) edges. Features referenced inside relations are shown in uppercase.}
\label{fig:graph-example}
\end{figure}

An \textbf{invention graph} represents a patent's technical content as two complementary structures (Figure~\ref{fig:graph-example}). The \textit{feature hierarchy} is a tree of technical components organized by part-of and example-of relationships, signaled by terms such as \textit{comprising} and \textit{including}. \textit{Functional relations} capture cross-cutting dependencies between features via verbal constructions (e.g., ``\textsc{Engine} \textit{configured to rotate} \textsc{Wheels}''). Features have a single hierarchical parent but may participate in multiple relations, yielding a directed acyclic graph (DAG). In the neural parser, both structures are modeled as per-word parent predictions forming tree-like hierarchies.

\paragraph{Task Definition.} The model operates at the word level, predicting for each word: (1)~a \textit{parent} in the feature hierarchy (or a pseudo-node for non-graph words), (2)~an \textit{edge label} distinguishing within-phrase (syntactic) from between-feature (semantic) edges, and (3)~a parent in the relation hierarchy. Syntactic edges are used in postprocessing to assemble multi-word feature spans from individual word predictions.

\paragraph{Architecture Choice.} A deployment constraint shapes the architecture: the neural parser must produce graphs structurally identical to rule-based output---same node types, edge labels, and relation format---for backward compatibility and controlled A/B testing in the search pipeline of~\cite{daniell2025efficient}. This rules out alternative approaches (e.g., NER with heuristic tree assembly, sequence-to-sequence generation) that would produce different graph formats. The per-word parent prediction formulation directly mirrors the rule-based system's output, making biaffine structured prediction~\cite{dozat2017deep} the natural choice. LLM-based extraction is prohibitively expensive at corpus scale---parsing 150 million documents costs over \$80{,}000 in input tokens alone\footnote{Estimated using Gemini 2 Flash Lite batch API (\$0.0375 per million input tokens)} while output tokens at least double this vs.\ under \$10{,}000 on spot GPU instances (Section~\ref{sec:efficiency})---and introduces rate-limit and schema-consistency risks. LLM-generated labels nonetheless remain promising for augmenting training data beyond the rule-based teacher ceiling.

\subsection{Model Architecture}
\label{sec:model-architecture}

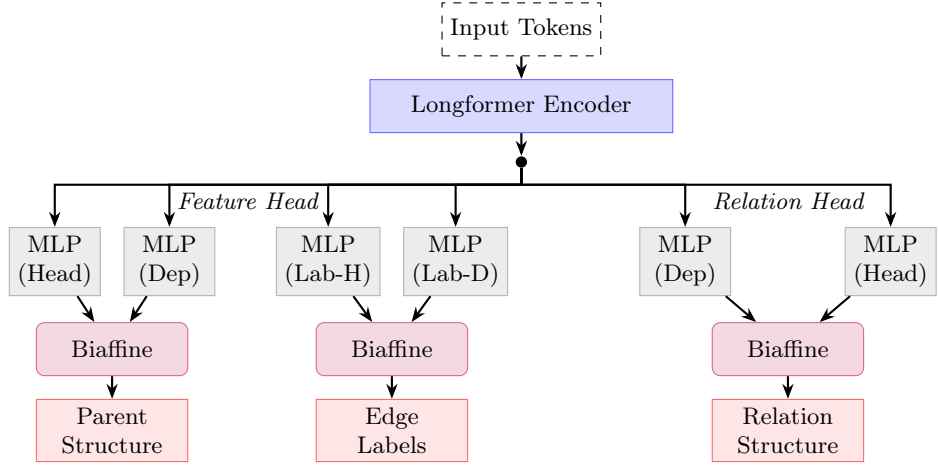
\begin{figure}[t]
\centering
\begin{tikzpicture}[
    node distance=0.8cm and 1.2cm,
    box/.style={rectangle, draw, minimum width=2cm, minimum height=0.7cm, align=center, font=\small},
    encoder/.style={box, fill=blue!15, draw=blue!60, minimum width=4cm},
    mlp/.style={box, fill=gray!15, draw=gray!60, minimum width=1.2cm, minimum height=0.5cm},
    biaffine/.style={box, fill=purple!15, draw=purple!60, rounded corners=3pt},
    output/.style={box, fill=red!10, draw=red!60, minimum width=2.0cm},
    arrow/.style={-{Stealth[length=2mm]}, thick},
    split/.style={circle, fill=black, inner sep=1.5pt}
]

\node[box, draw, dashed] (input) {Input Tokens};

\node[encoder, below=0.3cm of input] (encoder) {Longformer Encoder};

\node[split, below=0.3cm of encoder] (split) {};

\node[mlp, below left=0.8cm and 5.5cm of split] (arc_d) {MLP\\(Head)};
\node[mlp, right=0.3cm of arc_d] (arc_h) {MLP\\(Dep)};
\node[mlp, right=0.8cm of arc_h] (lab_d) {MLP\\(Lab-H)};
\node[mlp, right=0.3cm of lab_d] (lab_h) {MLP\\(Lab-D)};

\node[biaffine, below=0.8cm of $(arc_d)!0.5!(arc_h)$] (biaffine1) {Biaffine};
\node[biaffine, below=0.8cm of $(lab_d)!0.5!(lab_h)$] (biaffine3) {Biaffine};

\node[mlp, below right=0.8cm and 1.5cm of split] (rel_h) {MLP\\(Dep)};
\node[mlp, right=1.5cm of rel_h] (rel_d) {MLP\\(Head)};

\node[biaffine, below=0.8cm of $(rel_h)!0.5!(rel_d)$] (biaffine2) {Biaffine};

\node[output, below=0.3cm of biaffine1] (out1) {Parent\\Structure};
\node[output, below=0.3cm of biaffine3] (out2) {Edge\\Labels};
\node[output, below=0.3cm of biaffine2] (out3) {Relation\\Structure};

\draw[arrow] (input) -- (encoder);
\draw[arrow] (encoder) -- (split);

\draw[arrow] (split) -- ++(0,-0.3) -| (arc_d);
\draw[arrow] (split) -- ++(0,-0.3) -| (arc_h);
\draw[arrow] (split) -- ++(0,-0.3) -| (lab_d);
\draw[arrow] (split) -- ++(0,-0.3) -| (lab_h);
\draw[arrow] (split) -- ++(0,-0.3) -| (rel_h);
\draw[arrow] (split) -- ++(0,-0.3) -| (rel_d);

\draw[arrow] (arc_d) -- (biaffine1);
\draw[arrow] (arc_h) -- (biaffine1);
\draw[arrow] (lab_d) -- (biaffine3);
\draw[arrow] (lab_h) -- (biaffine3);
\draw[arrow] (rel_h) -- (biaffine2);
\draw[arrow] (rel_d) -- (biaffine2);

\draw[arrow] (biaffine1) -- (out1);
\draw[arrow] (biaffine3) -- (out2);
\draw[arrow] (biaffine2) -- (out3);

\node at ($(arc_d.north)!0.5!(lab_h.north) + (-0.1, 0.35)$) [font=\footnotesize\itshape] {Feature Head};
\node at ($(rel_h.north)!0.5!(rel_d.north) + (0, 0.35)$) [font=\footnotesize\itshape] {Relation Head};

\end{tikzpicture}
\caption{Model architecture. The Feature Head predicts parent structure and edge labels (syntactic vs.\ semantic); the Relation Head predicts functional relationships.}
\label{fig:architecture}
\end{figure}

Our architecture extends the biaffine dependency parser of Dozat and Manning~\cite{dozat2017deep} for invention graph induction. As shown in Figure~\ref{fig:architecture}, the model has three components: a contextual encoder and two parallel prediction heads.

\paragraph{Encoder.} We define \textit{words} as whitespace-delimited units of the input text. Each word is encoded into a fixed number of subword tokens ($k{=}3$) by the pretrained tokenizer, and the full subword sequence is processed by XLM-RoBERTa-Longformer~\cite{beltagy2020longformer}. A learned scalar mix over the last 4 hidden layers is followed by mean pooling over each word's subword tokens, producing one representation per word. All subsequent scoring operates at this word level. To extend the encoder beyond its native 16,384-subword limit, we tile the learned position embedding matrix to support sequences up to 65,536 subwords. This requires no retraining: sequences within the original range use pretrained positions unchanged; longer sequences receive repeated embeddings that are suboptimal but sufficient under local attention.

\paragraph{Biaffine Scoring.} Each prediction head projects encoder outputs through separate MLPs for dependent and head roles, then scores pairs using biaffine attention~\cite{dozat2017deep}:
\begin{equation}
s(i, j) = {\mathbf{h}_i^{\text{dep}}}^\top \mathbf{W} \mathbf{h}_j^{\text{head}} + \mathbf{b}^\top \mathbf{h}_i^{\text{dep}}
\label{eq:biaffine}
\end{equation}
The \textbf{Feature Head} uses two parallel biaffine scorers: one for parent prediction ($\mathbf{S}^{\text{arc}} \in \mathbb{R}^{n \times n}$) and one for edge label classification ($\mathbf{S}^{\text{label}} \in \mathbb{R}^{n \times n \times K}$), distinguishing syntactic edges (within-phrase) from semantic edges (between features). The \textbf{Relation Head} uses a single biaffine scorer ($\mathbf{S}^{\text{rel}} \in \mathbb{R}^{n \times n}$) for functional relationships. Unlike dependency parsing, where every word belongs to a single-rooted tree, invention graphs admit multiple roots and exclude most document words. Two pseudo-nodes---\textsc{None} (index~0) and \textsc{Root} (index~1)---model these decisions, allowing each word to be classified as a non-feature, a root feature, or attached to a parent within the graph.

\subsection{Local Biaffine Attention}
\label{sec:local-attention}

\begin{figure}[t]
\centering
\includegraphics[width=\linewidth]{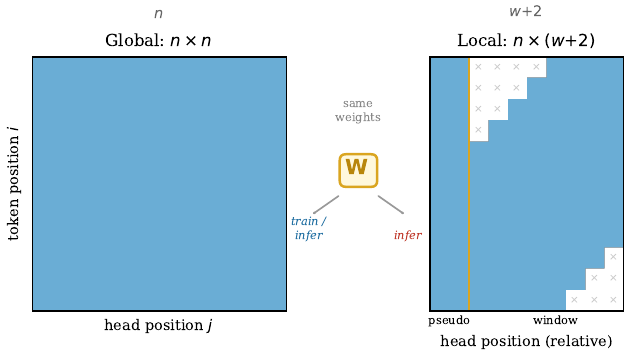}
\caption{Global vs.\ local biaffine score matrices. Global attention computes all $n^2$ scores; local attention computes only $n(w{+}2)$ scores within a sliding window plus two pseudo-node columns. Both modes share the same weight matrix~$\mathbf{W}$, enabling training with global attention on short sequences and inference with local attention on long sequences.}
\label{fig:attention-pattern}
\end{figure}

Standard biaffine attention computes all $O(n^2)$ pairwise scores, which is prohibitive for documents exceeding 40,000 words. Following the sliding-window principle of~\cite{wang2024docban}, our \textit{local biaffine attention} reduces complexity to $O(n \cdot w)$ by restricting each word to score only positions within a window of size $w$, plus two pseudo-nodes. For word $i$, the candidate set is:
\begin{equation}
\mathcal{N}(i) = \{\textsc{None}, \textsc{Root}\} \cup \{j : |i - j| \leq w/2\}
\end{equation}
This produces a score tensor of shape $[n, w{+}2]$ instead of $[n, n]$, as illustrated in Figure~\ref{fig:attention-pattern}.

\paragraph{Weight Sharing.} A key property of our local attention is that it uses \textit{exactly the same weight matrix}~$\mathbf{W}$ as global biaffine attention---only the set of positions scored differs. This enables training with global attention on short sequences (600 words) and deploying with local attention on long documents (40,000+ words), or switching window sizes at inference time without retraining.

\paragraph{Implementation.} We compute local scores efficiently using a chunking strategy adapted from Longformer~\cite{beltagy2020longformer}: the sequence is divided into overlapping chunks of size $2w$ with stride $w$, biaffine scores are computed within each chunk, and diagonal bands are extracted via a skew operation to produce the sliding window format. Pseudo-node scores are computed globally (as each word may be \textsc{None} or \textsc{Root} regardless of position) and concatenated with the window scores. Words near sequence boundaries have their windows masked to prevent attending to invalid positions.

\paragraph{Locality Justification.} Patent text exhibits strong locality in hierarchical relationships: a component is typically described near its parent in the claim structure. Analysis of rule-parsed training graphs confirms that the vast majority of parent-child edges fall within a small window (Section~\ref{sec:window-size}); the relation head similarly benefits, as functional relationships are usually expressed in contiguous text spans.

\subsubsection{Training Objective}
\label{sec:training-objective}

Both the feature parent and relation heads are trained with weighted cross-entropy:
\begin{equation}
\mathcal{L}_{\text{head}} = -\sum_{i \in \mathcal{M}} \alpha_{y_i} \log \frac{\exp(S_{i, y_i})}{\sum_j \exp(S_{i,j})}
\end{equation}
where $\mathcal{M}$ is the set of non-padded word positions and $\alpha_k$ are per-class weights. Since most words do not participate in the graph, the \textsc{None} class dominates; we set $\alpha_{\textsc{None}} = 0.05$ for the feature head and $0.1$ for the relation head, with all other positions receiving weight~$1.0$ (see ablation in Section~\ref{sec:ablation}). The edge label head uses standard cross-entropy at the gold parent position~\cite{dozat2017deep}. The total loss is $\mathcal{L} = \mathcal{L}_{\text{parent}} + \mathcal{L}_{\text{label}} + \mathcal{L}_{\text{rel}}$.

\subsection{Training Data and Setup}
\label{sec:training-data}

\subsubsection{Dataset Construction}
\label{sec:dataset-construction}

The training dataset is constructed using invention graphs generated by the rule-based system of~\cite{bjorkqvist2023building}, which has been deployed in production for over 5 years and has processed more than 150 million patent documents. This system uses spaCy~\cite{spacy2020} for linguistic analysis combined with regular expression patterns for relation extraction.

\paragraph{Training Set.} We sample 1 million patent documents from the full patent corpus, stratified by jurisdiction: 60\% US patents, 20\% Chinese patents (translated to English), and 20\% from other jurisdictions.

To prevent data leakage and near-duplicate contamination, we deduplicate by patent family ID: the same invention may appear under different document identifiers across jurisdictions (e.g., US, EP, CN) with near-identical text, so only one document per family is retained across all splits. All families appearing in either the intrinsic evaluation set (5,000 documents) or the downstream search evaluation set (90,082 queries) are excluded from training entirely.

To enable efficient training with global attention, we apply a maximum sequence length of 600 words. Documents are truncated to this limit, but we require a minimum coverage threshold of 70\%---documents where truncation would retain less than 70\% of the original graph structure are excluded.

\paragraph{Test Set.} We hold out 5,000 documents randomly sampled from the patent corpus for evaluation. Unlike training, test documents are processed at full length using local attention, enabling assessment of generalization to long documents not seen during training.

\subsubsection{Training Configuration}
\label{sec:training-config}

Table~\ref{tab:hyperparameters} summarizes our training hyperparameters.

\begin{table}[t]
\centering
\caption{Training hyperparameters. Architecture and regularization details in Sections~\ref{sec:model-architecture}--\ref{sec:training-objective}.}
\label{tab:hyperparameters}
\begin{tabular}{ll}
\toprule
\textbf{Parameter} & \textbf{Value} \\
\midrule
MLP dimensions (parent / label / relation) & 256 / 128 / 256 \\
Max subword tokens per word & 3 \\
Optimizer & AdamW (lr $= 10^{-4}$) \\
LR ratio (encoder vs.\ task layers) & 1:10 \\
Warmup & 10\% of total steps, then linear decay \\
Batch size (tokens per worker) $\times$ accumulation & 1{,}200 $\times$ 16 \\
GPUs (NVIDIA L4, data-parallel) & 8 \\
Feature \textsc{None} weight / Relation \textsc{None} weight & 0.05 / 0.1 \\
Encoder dropout / MLP shared dropout & 0.1 / 0.33 \\
Gradient clipping (max norm) & 2.0 \\
Precision & bfloat16 mixed \\
\bottomrule
\end{tabular}
\end{table}

Training runs for 100 epochs distributed across 8 NVIDIA L4 GPUs using Ray data-parallel, taking approximately 60 hours. We select the best checkpoint by whole graph IoU (Section~\ref{sec:intrinsic-metrics}) on the validation set.

\section{Experimental Setup}
\label{sec:experiments}

\subsection{Evaluation Metrics}
\label{sec:metrics}

\subsubsection{Intrinsic Graph Quality Metrics}
\label{sec:intrinsic-metrics}

We compare predicted graphs against rule-based teacher graphs by representing each graph as a set of \textit{(parent string, child string)} edge tuples and computing set-based precision, recall, and IoU (Intersection over Union). \textbf{Whole Graph IoU}, our primary metric, pools feature hierarchy and functional relation edges into a single set. We also report IoU at finer granularities (feature edges only, nodes, words), \textbf{IoU@$k$} (percentage of documents with IoU $\geq k/100$), and \textbf{coverage}---the fraction of document words participating in the graph, reported as total word coverage and separately for features and relations. A systematic bias affects string matching: the rule-based parser normalizes extracted text (e.g., lemmatization), while the neural parser is purely extractive, so semantically correct predictions may count as mismatches. Since no human-annotated corpus exists, intrinsic metrics measure teacher agreement; the extrinsic search evaluation serves as the primary quality signal.

\subsubsection{Extrinsic Search Evaluation}
\label{sec:extrinsic-metrics}

To evaluate the impact of graph quality on downstream search, the neural parser is integrated into the Graph Transformer retrieval system of~\cite{daniell2025efficient}, which is trained on examiner-cited prior art to rank candidate patents by relevance. On the evaluation set, this system with rule-based parser (Recall@3 = 0.416) substantially outperforms dense retrieval: PaECTER~\cite{ghosh2024paecter} achieves 0.280, Stella~\cite{zhang2024stella} 0.273, and GTE-ModernBERT~\cite{zhang2024gte} 0.200, despite using up to 4{,}096 tokens per chunk. The rule-based parser thus represents a strong baseline rather than a straw man.

\paragraph{Citation-Based Ground Truth.} We evaluate using examiner-cited references as relevance labels, considering both X citations (documents particularly relevant to novelty of the claimed invention) and Y citations (documents relevant when combined with other references, challenging inventive step). We report citation-averaged Recall@$k$:
\begin{equation}
\text{Recall@}k = \frac{1}{|Q|} \sum_{q \in Q} \frac{|\text{Retrieved}_k(q) \cap \text{Citations}(q)|}{|\text{Citations}(q)|}
\end{equation}
where $Q$ is the query set and $\text{Retrieved}_k(q)$ the top-$k$ candidates. We use $k{=}3$ as the primary metric. Results are reported on three subsets: all 90{,}082 queries, the 11{,}698 chemical patent queries (identified by CPC classification classes C01, C05, C07, C08, C09, C10, C11, and C13), and the remaining 78{,}384 non-chemical queries.

\paragraph{Intrinsic Evaluation.} We evaluate graph quality on the held-out test set of 5,000 documents described in Section~\ref{sec:dataset-construction}, processed at full length.

\section{Results}
\label{sec:results}

\subsection{Main Results}
\label{sec:main-results}

\subsubsection{Intrinsic Graph Quality}

Table~\ref{tab:intrinsic} compares the best neural parser against the rule-based teacher on the test set of 5,000 documents.

\begin{table}[t]
\centering
\caption{Intrinsic graph quality on the test set (5,000 documents). The rule-based teacher achieves IoU=1.0 by definition since it generates the training labels.}
\label{tab:intrinsic}
\begin{tabular}{lccccc}
\toprule
\textbf{Model} & \textbf{IoU} & \textbf{IoU@90} & \textbf{Precision} & \textbf{Recall} & \textbf{Coverage} \\
\midrule
Rule-based (teacher) & 1.000 & 1.000 & 1.000 & 1.000 & 0.847 \\
Neural & \textbf{0.653} & \textbf{0.142} & \textbf{0.761} & \textbf{0.780} & \textbf{0.861} \\
\bottomrule
\end{tabular}
\end{table}

The neural parser achieves 0.653 IoU with the rule-based teacher, measuring distillation fidelity rather than absolute quality. Word coverage exceeds the teacher (0.861 vs 0.847), indicating richer graph structures; whether this translates to search improvement is answered below (training details in Section~\ref{sec:ablation}).

\subsubsection{Search Recall}

Table~\ref{tab:search} presents extrinsic evaluation on the downstream search task using 90{,}082 queries, each paired with a known relevant reference retrieved from 168{,}809 documents. We evaluate two configurations: (1)~\textit{Claims}---only query graphs are re-parsed with the neural parser; (2)~\textit{Descriptions}---both query and corpus document graphs are re-parsed, using a 3-million-document neural-parsed subset (see Section~\ref{sec:efficiency}).

\begin{table}[t]
\centering
\caption{Search recall (Recall@3). \textit{Claims}: queries re-parsed, corpus rule-based. \textit{Descriptions}: both re-parsed on a corpus subset. Significance via McNemar's paired test (one-sided): $^{*}p < 0.05$, $^{**}p < 0.01$.}
\label{tab:search}
\begin{tabular}{llccc}
\toprule
\textbf{Configuration} & \textbf{Parser} & \textbf{All} & \textbf{Chemical} & \textbf{Non-Chemical} \\
\midrule
\multirow{2}{*}{Claims} & Rule-based & 0.4163 & 0.3790 & 0.4218 \\
 & Neural & \textbf{0.4183}$^{*}$ & \textbf{0.3799} & \textbf{0.4240}$^{*}$ \\
\midrule
\multirow{2}{*}{Desc.} & Rule-based & 0.3336 & 0.3060 & 0.3377 \\
 & Neural & \textbf{0.3373}$^{**}$ & \textbf{0.3095} & \textbf{0.3414}$^{**}$ \\
\bottomrule
\end{tabular}
\end{table}

Both overall improvements are statistically significant under McNemar's paired test over 90{,}082 queries ($p = 0.04$ for claims; $p < 0.001$ for descriptions). Re-parsing both queries and corpus documents yields the larger gain (+1.1\% vs +0.5\%), confirming that consistent neural graphs on both sides of retrieval---rather than query-side improvements alone---are the key driver.

The chemical subset does not reach significance ($p = 0.42$ claims, $p = 0.14$ descriptions), despite showing the largest qualitative improvements in graph structure (Figure~\ref{fig:comparison}). This suggests the chemical validation set may be too small to capture observed gains; domain-specific evaluation protocols remain future work.

\subsubsection{Graph Structure Analysis}

Table~\ref{tab:structure} compares structural properties of graphs generated by each parser.

\begin{table}[t]
\centering
\caption{Graph structure comparison between rule-based and neural parsers.}
\label{tab:structure}
\begin{tabular}{lcc}
\toprule
\textbf{Metric} & \textbf{Rule-based} & \textbf{Neural} \\
\midrule
Graph Depth & 2.157 & 2.057 \\
Word Coverage & 0.847 & 0.862 \\
Feature Coverage & 0.514 & 0.521 \\
Relation Coverage & 0.369 & 0.376 \\
\bottomrule
\end{tabular}
\end{table}

\begin{figure}[t]
\centering
\includegraphics[width=\linewidth]{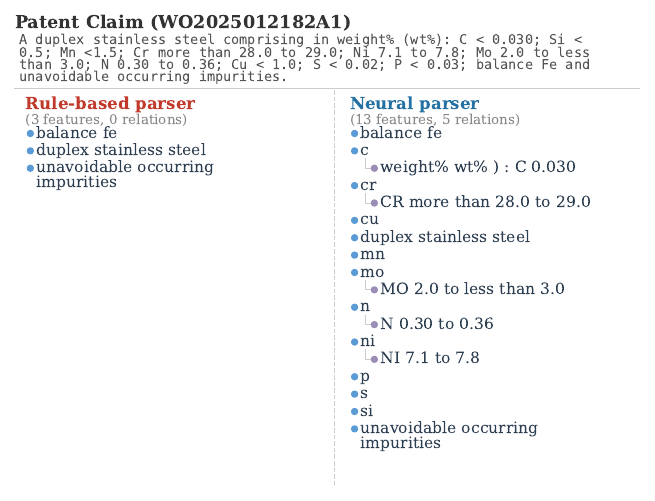}
\caption{Rule-based vs.\ neural parser output for a chemical patent claim (WO2025012182A1). The rule-based parser extracts only 3 features, while the neural parser identifies 13 features and 5 relations capturing element compositions. This domain---chemical compositions with numerical ranges---is particularly challenging for hand-crafted rules.}
\label{fig:comparison}
\end{figure}

The neural parser produces slightly shallower graphs (depth 2.057 vs 2.157) while achieving higher coverage across all metrics, suggesting a more balanced representation. Figure~\ref{fig:comparison} illustrates this on a chemical patent: the neural parser extracts 13 chemical constituents as individual features where the rule-based parser captures only 3 top-level features, directly benefiting downstream search for specific alloy compositions.

\subsection{Ablation Studies}
\label{sec:ablations}

\subsubsection{Training Configuration Ablation}
\label{sec:ablation}

Table~\ref{tab:ablation} summarizes the effect of key training decisions. We vary training set size, document length filtering, loss weights, and root penalization. All models are evaluated at the best checkpoint (maximum IoU over 100 epochs) on the same 5{,}000-document test set.

\begin{table}[t]
\centering
\caption{Ablation of training configuration. All models use 100 training epochs; metrics reported at best checkpoint (max IoU). Coverage: fraction of document words assigned to graph features.}
\label{tab:ablation}
\begin{tabular}{lcccc}
\toprule
\textbf{Configuration} & \textbf{IoU} & \textbf{P} & \textbf{R} & \textbf{Cov.} \\
\midrule
100K dataset & 0.614 & 0.727 & 0.753 & 0.797 \\
1M dataset & 0.651 & 0.766 & 0.772 & 0.846 \\
\quad + short-doc filter, loss tuning & \textbf{0.653} & \textbf{0.761} & \textbf{0.780} & 0.861 \\
\quad + root penalization & 0.637 & 0.748 & 0.768 & \textbf{0.866} \\
\bottomrule
\end{tabular}
\end{table}

\paragraph{Data scale.} Scaling from 100K to 1M training examples yields the largest single improvement: IoU rises from 0.614 to 0.651, with coverage increasing from 0.797 to 0.846 as the model generalizes to a wider vocabulary of feature expressions.

\paragraph{Document length filtering.} Training exclusively on documents with $\leq$600 words (combined with the 70\% coverage filter from Section~\ref{sec:dataset-construction}) improves recall from 0.772 to 0.780 while maintaining IoU, even though inference uses full-length documents. The rule-based teacher produces higher-quality labels on shorter patents, and the model learns structural patterns that generalize beyond the training length distribution.

\paragraph{Loss weights and root penalization.} Down-weighting the \textsc{None} class is essential: asymmetric weights (feature 0.05, relation 0.10) outperform symmetric alternatives. Increasing the root loss weight to encourage deeper structures does not improve IoU (0.637 vs 0.653) and decreases precision; we use the default equal root weight.

\subsubsection{Window Size and Document Length}
\label{sec:window-size}

Analysis of the training data confirms that 98\% of parent-child relationships occur within 2048 words, justifying local attention. To evaluate window size at inference time, we measure search recall using all claims combined as queries (concatenation of all independent claims per patent), since global attention on full corpus documents exceeds memory.

\begin{table}[t]
\centering
\caption{Effect of attention window size on search recall (Recall@3), measured using all-claims queries. In production, an adaptive strategy selects the largest window fitting NVIDIA L4 memory (24\,GB): global for $<$5K words, 2048 for 5K--15K, 1024 for $>$15K.}
\label{tab:window-size}
\begin{tabular}{lcc}
\toprule
\textbf{Window Size} & \textbf{Recall@3} & \textbf{Doc.\ Range} \\
\midrule
512 & 0.4658 & --- \\
1024 & 0.4676 & $>$15K words \\
2048 & 0.4690 & 5K--15K words \\
Global & 0.4714 & $<$5K words \\
\bottomrule
\end{tabular}
\end{table}

Larger windows consistently improve recall (Table~\ref{tab:window-size}), with global attention as the upper bound. The adaptive strategy maximizes recall within the GPU memory budget by assigning each document the largest feasible window.

\subsection{Deployment Experience}
\label{sec:efficiency}

The search pipeline has two parsing stages: \textit{query parsing} at search time and \textit{corpus indexing} of full patent descriptions offline. The neural parser is deployed in production for query parsing, replacing the rule-based parser with no user-facing disruptions; an internal evaluation by domain experts with patent search background confirmed structural equivalence for most documents, with neural graphs capturing additional features in challenging domains such as chemical compositions (Figure~\ref{fig:comparison}). This corresponds to the Claims configuration (+0.5\% recall). For corpus indexing, we have validated inference quality and cost on a 3-million-document subset (the Descriptions configuration, +1.1\% recall); scaling to the full 150 million documents remains in progress.

\paragraph{Online query inference.} The biaffine model runs on fourth-generation Intel Xeon Scalable processors with native \texttt{bfloat16} support, eliminating the need for GPUs in online serving. Throughput increases from 2 to 10 queries per second per CPU instance, and average latency drops from 2--3 seconds to under 1 second. Long queries that previously timed out after 60+ seconds (common for chemical patents with complex formulas) complete in under 5 seconds; inputs are truncated at 40{,}000 characters as a safety bound.

\paragraph{Batch corpus parsing.} We parsed 3 million documents as a single distributed Ray Data job on 10 NVIDIA L4 spot GPU instances, with CPUs on the same nodes handling pre- and post-processing. The job completed in two days without out-of-memory errors. GPU utilization exceeded 70\%, though variable-size tree outputs create packing challenges in Ray Data workers that limit further utilization. At current utilization, batch parsing cost is estimated at 3$\times$ lower than the rule-based system; improving GPU utilization to 90\%+ would bring this to 5$\times$. Scaling beyond 3 million documents will require a full orchestration framework with improved observability and error recovery.

\paragraph{Edge cases and lessons learned.} The neural parser largely reproduces the rule-based system's strengths and weaknesses: domains where rules worked well remain strong, and challenging cases (e.g., Markush structures) remain difficult. One unexpected issue arose from training on short documents: the model preserved patent jargon such as ``according to claim~1'' that the rule-based parser had filtered out, especially in longer inputs combining claims and descriptions. Rather than retraining on longer sequences---which risked degrading metrics (Section~\ref{sec:ablation})---we applied a simple regex post-filter.

\section{Conclusion and Future Work}
\label{sec:conclusion}

We showed that biaffine structured prediction, made tractable for long documents via local attention with shared weights, can replace brittle rule-based patent graph extraction. Knowledge distillation from 1 million rule-parsed examples produces a model that surpasses its teacher on downstream citation recall across both query-only and full-pipeline configurations, while reducing inference cost. Future work includes expert-annotated data to move beyond the teacher ceiling, joint training with the search model, and multilingual extension.

\paragraph{Use of Generative AI.} Generative AI tools (Claude, Anthropic) were used to assist with proofreading and improving readability of the manuscript, and to generate code for producing figures. All scientific content, experimental design, and conclusions are solely the work of the authors.

\bibliographystyle{splncs04}
\bibliography{references}

\begin{thebibliography}{10}
\providecommand{\url}[1]{\texttt{#1}}
\providecommand{\urlprefix}{URL }
\providecommand{\doi}[1]{https://doi.org/#1}

\bibitem{acikalin2022patent}
Acikalin, U.U., Kutlu, M.: Patent search using triplet networks based
  fine-tuned {SciBERT}. arXiv preprint arXiv:2207.11497  (2022)

\bibitem{beltagy2020longformer}
Beltagy, I., Peters, M.E., Cohan, A.: Longformer: The long-document
  transformer. arXiv preprint arXiv:2004.05150  (2020)

\bibitem{bjorkqvist2023building}
Bj{\"o}rkqvist, S., Kallio, J.: Building a graph-based patent search engine.
  In: Proceedings of the 46th International ACM SIGIR Conference on Research
  and Development in Information Retrieval. pp. 3300--3304 (2023).
  \doi{10.1145/3539618.3591842}

\bibitem{clarke2018basics}
Clarke, N.S.: The basics of patent searching. World Patent Information
  \textbf{54},  S4--S10 (2018). \doi{10.1016/j.wpi.2017.02.006}

\bibitem{daniell2025efficient}
Daniell, K., Buzhinsky, I., Bj{\"o}rkqvist, S.: Efficient patent searching
  using graph transformers. In: PatentSemTech'25: 6th Workshop on Patent Text
  Mining and Semantic Technologies (2025)

\bibitem{dhondt2010clef}
D'hondt, E., Verberne, S.: {CLEF-IP} 2010: Prior art retrieval using the
  different sections in patent documents. In: Proceedings of CLEF 2010 (2010)

\bibitem{dozat2017deep}
Dozat, T., Manning, C.D.: Deep biaffine attention for neural dependency
  parsing. In: Proceedings of the 5th International Conference on Learning
  Representations (ICLR) (2017)

\bibitem{dwivedi2021generalization}
Dwivedi, V.P., Bresson, X.: A generalization of transformer networks to graphs.
  arXiv preprint arXiv:2012.09699  (2021)

\bibitem{freunek2021bert}
Freunek, M., Bodmer, A.: {BERT} based patent novelty search by training claims
  to their own description. arXiv preprint arXiv:2103.01126  (2021)

\bibitem{geng2021open}
Geng, Z., Chen, G., Han, Y., Lu, G., Li, F.: Open relation extraction from
  patent claims using hybrid neural networks. Expert Systems with Applications
  \textbf{170},  114482 (2021)

\bibitem{ghosh2024paecter}
Ghosh, M., Erhardt, S., Rose, M.E., Buunk, E., Harhoff, D.: {PaECTER}:
  Patent-level representation learning using citation-informed transformers.
  arXiv preprint arXiv:2402.19411  (2024)

\bibitem{hinton2015distilling}
Hinton, G., Vinyals, O., Dean, J.: Distilling the knowledge in a neural
  network. arXiv preprint arXiv:1503.02531  (2015)

\bibitem{spacy2020}
Honnibal, M., Montani, I., Van~Landeghem, S., Boyd, A.: {spaCy}:
  Industrial-strength natural language processing in {Python} (2020),
  \url{https://spacy.io}

\bibitem{merrouni2025patenteb}
Merrouni, I.A., Belkadi, A., Nehar, A., Boumhidi, I.: {PatenTEB}: A
  comprehensive benchmark and model family for patent text embedding. arXiv
  preprint arXiv:2510.22264  (2025)

\bibitem{reimers2019sentence}
Reimers, N., Gurevych, I.: Sentence-{BERT}: Sentence embeddings using siamese
  {BERT}-networks. In: Proceedings of the 2019 Conference on Empirical Methods
  in Natural Language Processing and the 9th International Joint Conference on
  Natural Language Processing. pp. 3982--3992 (2019)

\bibitem{tai2015improved}
Tai, K.S., Socher, R., Manning, C.D.: Improved semantic representations from
  tree-structured long short-term memory networks. In: Proceedings of the 53rd
  Annual Meeting of the Association for Computational Linguistics and the 7th
  International Joint Conference on Natural Language Processing (Volume 1: Long
  Papers). pp. 1556--1566 (2015). \doi{10.3115/v1/P15-1150}

\bibitem{wang2024docban}
Wang, X., Shen, J., Li, J., Liu, Y.: {DocBAN}: An efficient biaffine attention
  network for document-level named entity recognition. In: Advanced Intelligent
  Computing Technology and Applications (ICIC 2024). pp. 65--76. Springer
  (2024). \doi{10.1007/978-981-97-5669-8_6}

\bibitem{ying2021transformers}
Ying, C., Cai, T., Luo, S., Zheng, S., Ke, G., He, D., Shen, Y., Liu, T.Y.: Do
  transformers really perform bad for graph representation? arXiv preprint
  arXiv:2106.05234  (2021)

\bibitem{zaheer2020bigbird}
Zaheer, M., Guruganesh, G., Dubey, A., Ainslie, J., Alberti, C., Ontanon, S.,
  Pham, P., Ravula, A., Wang, Q., Yang, L., Ahmed, A.: {Big Bird}: Transformers
  for longer sequences. In: Advances in Neural Information Processing Systems.
  vol.~33, pp. 17283--17297 (2020)

\bibitem{zhang2024stella}
Zhang, D.: Jasper and stella: distillation of sota embedding models. Technical
  report, NovaSearch  (2024),
  \url{https://huggingface.co/NovaSearch/stella_en_400M_v5}

\bibitem{zhang2024gte}
Zhang, X., Li, Y., Zhang, W., Long, D., Xie, P.: {GTE-ModernBERT}: Modernized
  bi-directional encoder for general text embeddings. arXiv preprint
  arXiv:2412.13663  (2024)

\bibitem{zhu2024longembed}
Zhu, D., Wang, L., Yang, N., Song, Y., Wu, W., Wei, F., Li, S.: {L}ong{E}mbed:
  Extending embedding models for long context retrieval. In: Proceedings of the
  2024 Conference on Empirical Methods in Natural Language Processing. pp.
  802--816 (2024), \url{https://aclanthology.org/2024.emnlp-main.47/}

\end{thebibliography}

\end{document}